\documentclass[aps,prd,twocolumn,superscriptaddress,showpacs,showkeys,
nofootinbib]{revtex4-1}

\usepackage{latexsym}
\usepackage{amsmath}
\usepackage{amssymb}
\usepackage{amsfonts}
\usepackage{multirow}
\usepackage{xcolor}

\usepackage{supertabular} 
\usepackage{placeins}
\usepackage{epsfig}
\usepackage{graphicx}

\begin{document}

\title{Strange hidden-charm $P_{\psi s}^\Lambda(4459)$ and $P_{\psi s}^\Lambda(4338)$ pentaquarks and additional $P_{\psi s}^\Lambda$, $P_{\psi s}^\Sigma$ and $P_{\psi ss}^N$ candidates in a quark model approach }

\author{Pablo G. Ortega}
\email[]{pgortega@usal.es}
\affiliation{Departamento de Física Fundamental, Universidad de Salamanca, E-37008 Salamanca, Spain}
\affiliation{Instituto Universitario de F\'isica
Fundamental y Matem\'aticas (IUFFyM), Universidad de Salamanca, E-37008 Salamanca, Spain}

\author{David R. Entem}
\email[]{entem@usal.es}
\affiliation{Instituto Universitario de F\'isica
Fundamental y Matem\'aticas (IUFFyM), Universidad de Salamanca, E-37008 Salamanca, Spain}
\affiliation{Grupo de F\'isica Nuclear, Universidad de Salamanca, E-37008 Salamanca, Spain}

\author{Francisco Fern\'andez}
\email[]{fdz@usal.es}
\affiliation{Instituto Universitario de F\'isica
Fundamental y Matem\'aticas (IUFFyM), Universidad de Salamanca, E-37008 Salamanca, Spain}
\affiliation{Grupo de F\'isica Nuclear, Universidad de Salamanca, E-37008
Salamanca, Spain}

\date{\today}

\begin{abstract}

Hidden-charm pentaquark-like $P_{\psi s}^\Lambda(4459)^0$ and $P_{\psi s}^\Lambda(4338)$ resonances are studied in a constituent quark model as molecular meson-baryon structures.
Such states are found in the $J^P(I)=\frac{1}{2}^-(0)$ channel with masses and widths compatible with the experimental measurements in a coupled-channels calculation with all the parameters constrained from previous studies.
Other candidates are explored in the $J^P=\frac{1}{2}^-$, $\frac{3}{2}^-$ and $\frac{5}{2}^-$ channels in the charm and bottom sectors, with isospins $0$ ($P_{\psi s}^\Lambda$ and $P_{\Upsilon s}^\Lambda$) and $1$ ($P_{\psi s}^\Sigma$ and $P_{\Upsilon s}^\Sigma$).
Additionally, the formalism is extended to study the $P_{\psi ss}^N$ ($P_{\Upsilon ss}^N$) pentaquarks, where eight candidates are predicted as $\bar D_s \Xi_c$ molecules in $I=\frac{1}{2}$, with $J^P=\frac{1}{2}^-$, $\frac{3}{2}^-$ and $\frac{5}{2}^-$  for the charm sector and nine candidates as $B_s \Xi_b$ for the bottom one.
\end{abstract}



\maketitle


\section{Introduction}

The discovery of pentaquark states by the LHCb~\cite{LHCb:2015yax} revolutionized Hadron Physics, expanding the usual $qqq$ structure to four quarks and an antiquark.
The first observations detected in the $J/\psi p$ mass spectrum showed two resonances, dubbed $P_c(4380)^+$ and $P_c(4450)^+$, close to $D^{(*)}N$ thresholds, which
suggested a baryon-meson molecular nature in contrast to a compact pentaquark core. A further analysis showed that the higher resonance actually consisted in two separated narrow states, the $P_c(4440)^+$ and $P_c(4457)^+$, whereas a new state $P_c(4312)^+$ was detected~\cite{LHCb:2019kea}.

The detection of such pentaquarks, with minimum $\bar c cuud$ quark content, encouraged many authors~\cite{PhysRevLett.105.232001,Chen:2016ryt,Santopinto:2016pkp,Xiao:2019gjd} to predict similar hidden-charm structures with strangeness, i.e. with $\bar c c uds$, whose existence were recently confirmed with the discovery of the so-called $P_{cs}(4459)^0$~\cite{LHCb:2020jpq}.

The $P_{\psi s}^\Lambda(4459)^0$, following the nomenclature for exotic states proposed by LHCb Collaboration~\cite{Gershon:2022xnn}, was first spotted in the $J/\psi\Lambda$ invariant mass distribution from an analysis of the $\Xi_b^-\to J/\psi\Lambda K^-$ decays~\cite{LHCb:2020jpq} with a $3\sigma$ significance. The state is an isospin-0 pentaquark-like resonance with a $\bar ccuds$ minimum quark content.
The mass and width of this exotic state were measured to be

\begin{align}
 M_{P_{\psi s}^\Lambda(4459)^0} &= 4458.8\pm 2.9^{+4.7}_{-1.1}\,{\rm MeV/c}^2 \nonumber,\\
 \Gamma_{P_{\psi s}^\Lambda(4459)^0} &= 17.3\pm 6.5^{+8.0}_{-5.7}\,{\rm MeV}
\end{align}
that is, just $19$ MeV below the $\bar D^{*0}\Xi_c^0$.

Since its discovery, there has been a plethora of studies trying to unveil the nature of such resonance, among which the most popular explanation is the meson-baryon molecular structure~\cite{Chen:2020uif,Peng:2020hql,PhysRevD.103.034003,Chen:2020kco,Karliner:2022erb,Hofmann:2005sw,Anisovich:2015zqa,Feijoo:2015kts,Lu:2016roh,Zhang:2020cdi,Hu:2021nvs,Xiao:2021rgp,Zhu:2021lhd,Weng:2019ynv,Wang:2022gfb}.
The $J^P$ of the resonance was not experimentally determined due to a limited signal yield, but most studies favour a $J=\frac{1}{2}$ and/or $\frac{3}{2}$ with negative parity, which would allow the closest meson-baryon thresholds to be in a relative S-wave. Some studies (see e.g. Ref.~\cite{Xiao:2019gjd}), predicts two $J^P=\frac{1}{2}^-$ and $J^P=\frac{3}{2}^-$ resonances close in mass, which can be unresolved in the original $P_{\psi s}^\Lambda(4459)$ signal. The LHCb analyzed this two-resonance hypothesis, but could not confirm nor refute it.

Recently~\cite{LHCbtalk}, the LHCb has announced another $P_{\psi s}^\Lambda$ in the $J^P=\frac{1}{2}^-$ sector in the $B^-\to J/\psi\Lambda \bar p$ reaction, close to the $D^-\Xi_c^+$ threshold in S-wave with isospin 0, denoted $P_{\psi s}^\Lambda(4338)$, with mass and width:

\begin{align}
 M_{P_{\psi s}^\Lambda(4338)^0} &= 4338.2\pm 0.7\pm 0.4\,{\rm MeV/c}^2 \nonumber,\\
 \Gamma_{P_{\psi s}^\Lambda(4338)^0} &= 7.0\pm 1.2\pm 1.3\,{\rm MeV}
\end{align}
whose molecular nature has been explored in, e.g., Refs.~\cite{Yan:2022wuz,Wang:2022mxy,Karliner:2022erb,Wang:2022neq}.
This new discovery points to a rich spectroscopy of pentaquark-like states in the hidden-charm sector, whose exploration has just started.

In this work, we will explore the hidden-charm strange pentaquark states $P_{\psi s}^\Lambda$ as $\bar D^{(*)}\Xi_c^{(\prime)(*)}$ molecular states, and their bottom partners, in a coupled-channels formalism, using a constituent quark model which has been extensively used to describe meson and baryon spectrum~\cite{Vijande:2004he,Garcilazo:2001md,Segovia:2013wma} and, in particular, exotic states in the baryon spectrum as meson-baryon molecules~\cite{Ortega:2012cx,Ortega:2016syt,Ortega:2014fha}.

The paper is organized as follows: In Section~\ref{sec:model} we describe the details of the constituent quark model and the calculation of the resonances in a coupled-channels approach. In Sec.~\ref{sec:results} we present the results and in Sec.~\ref{sec:summary} we give a short summary.

\section{The model}\label{sec:model}

The QCD Lagrangian has a large global symmetry under $U(n_f)\times U(n_f)$ \emph{chiral} rotations of $n_f$ massless quark flavours. However, if chiral symmetry were conserved, we should see this symmetry in the
hadron spectra. For example, chiral symmetry would imply the existence of a partner of opposite parity for each meson, which is clearly not the case in the experimental meson spectra.
Current quark masses are non zero but have very small values that cannot explain the breaking in the spectra, therefore one can assume that chiral
symmetry is spontaneously broken by the QCD vacuum. Then, due to Goldstone theorem, light pseudo-Goldstone bosons emerge, mediating the interaction among quarks.
This phenomenology can be modeled using the instanton liquid mode of Ref.~\cite{Diakonov:2002fq}, which
assumes that quarks interact with fermionic zero modes of individual instantons, acquiring a dynamical mass.
A chiral invariant Lagrangian for quarks and Goldstone bosons which describes this effect is given by

\begin{equation}\label{eq:diakon}
 \mathcal{L}=\bar{\psi}(i\gamma^\mu \partial_\mu-M U^{\gamma_5})\psi
\end{equation}
where $U^{\gamma_5}=exp(i\phi^a\lambda^a\gamma_5/f_\pi)$, $\phi^a$ denotes the
pseudoscalar fields $(\vec\pi,K_i,\eta_8)$ with $i=(1,\ldots,4)$, $\lambda^a$
the $SU(3)$ flavor matrices and
$M$ the constituent quark mass.
The constituent quark masses are momentum-dependent. Their dependence can be directly derived from the underlying theory, but a convenient parametrization can be used as
$M(q^2)=m_qF(q^2)$, being $m_q\sim 300$ MeV and

\begin{equation}
 F(q^2)=\sqrt{\frac{\Lambda_\chi^2}{\Lambda_\chi^2+q^2}},
\end{equation}
where $\Lambda_\chi$ is a cut-off that controls the chiral symmetry-breaking scale.

Boson exchanges between light quarks show up when we expand the Goldstone boson field matrix $U^{\gamma^5}$
from Eq.~\eqref{eq:diakon} as

\begin{equation}
U^{\gamma_5}=1+\frac{i}{f_\pi}\gamma_5\lambda^a\phi^a-\frac{1}{2f_\pi^2}\phi^a\phi^a+\ldots\,.
\end{equation}
The first term can be identified with the constituent quark mass contribution, the second term
with the one-boson exchange, that is $\pi$, $K$ or $\eta$ exchanges, whereas the main contribution of the
third term can be modelled as a scalar $\sigma$ exchange.
Explicit expressions of the potentials derived from this expansion are detailed in Refs.~\cite{Vijande:2004he,Segovia:2008zz}.

For the heavy quark sector, where the chiral symmetry is explicitly broken, the dynamics is not governed by Goldstone bosons but QCD perturbative effects. Those are considered through the one-gluon exchange term derived from the Lagrangian,

\begin{equation}
\label{Lg}
{\mathcal L}_{gqq}=
i{\sqrt{4\pi\alpha _{s}}}\, \overline{\psi }\gamma _{\mu }G^{\mu
}_c \lambda _{c}\psi  \, ,
\end{equation}
where $\lambda _{c}$ are the SU(3) color generators and G$^{\mu }_c$ is the
gluon field. The strong coupling constant $\alpha_s$ has a scale dependence which allows to consistently describe light, strange and heavy mesons, whose explicit expression is,

\begin{align} \label{eq:alphascale}
 \alpha_s(\mu) = \frac{\alpha_0}{\ln\left(\frac{\mu^2+\mu_0^2}{\Lambda_0^2}\right)}
\end{align}
where $\mu$ is the reduced mass of the $q\bar q$ system and $\alpha_0$, $\mu_0$ and $\Lambda_0$ are parameters of the model.

The last piece is confinement, a QCD non-perturbative effect
which excludes colored hadrons in nature.
Within our CQM, this term is simulated by a linear-screened potential.
In this picture, a quark-antiquark can be viewed as linked by a one-dimensional color flux-tube.
At some scale, the spontaneous creation of a light $q\bar q$ pair can break the color flux-tube, which
saturates the potential at the same interquark distance. The potential is, then,
\begin{equation}
V_{CON}(\vec{r}_{ij})=\{-a_{c}\,(1-e^{-\mu_c\,r_{ij}})+ \Delta\}(\vec{%
\lambda^c}_{i}\cdot \vec{ \lambda^c}_{j})\,
\end{equation}
which behaves linearly at low $r$, with a strength given by $a_c\mu_c(\vec{\lambda^c}_{i}\cdot \vec{ \lambda^c}_{j})$, and a constant plateau at large $r$, given
by $(\Delta-a_c)(\vec{\lambda^c}_{i}\cdot \vec{ \lambda^c}_{j})$

 All the parameters of the model are constrained from previous studies on hadron phenomenology so, in this sense, the calculation is parameter free. This allows us to make robust predictions on the existence or nonexistence of specific molecular configurations. A first fit of the model parameters was done in Ref.~\cite{Vijande:2004he}, which analyzed the meson spectra from the aforementioned constituent quark model, previously used to study the $NN$ phenomenology and baryons spectrum. An additional small update of the original parameters was, then, done in Ref.~\cite{Segovia:2008zz} from the study of $J^{PC}=1^{--}$ charmonium spectroscopy.

\begin{figure}
 \includegraphics[width=.5\textwidth]{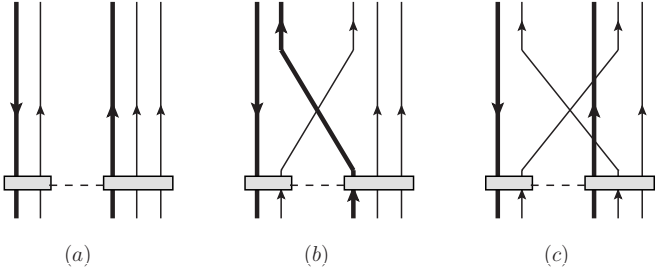}
 \caption{\label{fig:diagrams} Scheme of diagrams that are included in this work: Panel $(a)$ are direct diagrams, including the $\{\sigma,\pi,\eta\}$ exchanges in the $\bar D^{(*)}\Xi_c^{(\prime)(*)}\to \bar D^{(*)}\Xi_c^{(\prime)(*)}$ channels, $\{\sigma,\eta\}$ in $\bar D_s^*\Lambda_c \to \bar D_s^*\Lambda_c$ reactions and the $K$ exchange in $\bar D^{(*)}\Xi_c^{(\prime)(*)}\leftrightarrow \bar D_s^*\Lambda_c$ one. Panel $(b)$ are diagrams with a $c\leftrightarrow n$ quark exchange between clusters, which describe the $\bar D^{(*)}\Xi_c^{(\prime)(*)}\leftrightarrow J/\psi (\eta_c)\Lambda$. Panel $(c)$ are exchange diagrams which include a $n\leftrightarrow n$ exchange in $\bar D^{(*)}\Xi_c^{(\prime)(*)}\to \bar D^{(*)}\Xi_c^{(\prime)(*)}$ and $n \leftrightarrow s$ exchange in $\bar D^{(*)}\Xi_c^{(\prime)(*)}\leftrightarrow \bar D_s^*\Lambda_c$. The gray band represents the sum of interactions between quarks of different clusters, as detailed in the text (see Eqs.~\eqref{eq:directV},~\eqref{eq:exchangeHN} and~\eqref{eq:exchangeV}). The position of the charm quark is shown as a thick line. }
\end{figure}

The meson-baryon interaction emerges from the microscopic description at $qq$ level using the
Resonating Group Method (RGM). This way, we obtain an effective cluster-cluster interaction from the
underlying quark-quark dynamics, where the wave functions of the meson and baryon act as natural
cut-offs of the CQM potentials.

For the system under consideration, as we have identical quarks between clusters, we have to take into account the full antisymmetric wave function of the $\bar D^{(*)}\Xi_c^{(\prime)(*)}$.
 We can separate the full antisymmetric operator of the 3-identical quark system, up to a normalization factor, as ${\cal A}=1-P_{12}-P_{13}$, where $P_{12}(P_{13})$ is the operator that exchanges the $n$ quark from meson $\bar D^{(*)}$ (quark 1) and the second (third) $n$ quark from baryon $\Xi_c^{(\prime)(*)}$. That is, a sum of interactions that includes direct potentials, with no quark rearrangement among clusters, and exchange potentials, which do allow quark shifts among them. These contributions are schematically shown in Fig.~\ref{fig:diagrams}.

Then, the direct potentials can be written as

\begin{align}\label{eq:directV}
&
{}^{\rm RGM}V_{D}(\vec{P}',\vec{P}_{i}) = \sum_{i\in M, j\in B} \int d\vec{p}_{M'} d\vec{p}_{B'} d\vec{p}_{M} d\vec{p}_{B} \times \nonumber \\
&
\times \phi_{M}^{\ast}(\vec{p}_{M'}) \phi_{B}^{\ast}(\vec{p}_{B'})
V_{ij}(\vec{P}',\vec{P}_{i}) \phi_{M'}(\vec{p}_{M}) \phi_{B'}(\vec{p}_{B})  \,.
\end{align}
with $\vec P^{(\prime)}$ the initial (final) relative momentum of the meson-baryon (MB) system, $\vec p_{M(B)}$ the relative internal momentum of the meson (baryon) and $i$ $(j)$ the indices that run inside the meson (baryon) constituents.

The exchange diagrams, on the contrary, have two effects. On the one side, they add self-interaction to the $\bar D^{(*)}\Xi_c^{(\prime)(*)}$ channels if light quarks are exchanged and, on the other side, they connect different channels otherwise disconnected, such as $\bar D^{(*)}\Xi_c$ and $J/\psi\Lambda$ channels.

For the first case, the exchange kernel $^{\rm RGM}K$ can be written as,

\begin{align}
 {}^{\rm RGM}K(\vec P',\vec P_i) &= {}^{\rm RGM}H_E(\vec P',\vec P_i) - E_T\,{}^{\rm RGM} N_E(\vec P',\vec P_i)\,,
\end{align}
which is a non-local and energy-dependent kernel that we split in a potential term plus a normalization term, being $E_T$ the total energy of the system and $\vec P_i$ a continuous parameter.
Then, exchange Hamiltonian and normalization can be written as

\begin{align}\label{eq:exchangeHN}
&
{}^{\rm RGM}H_{E}(\vec{P}',\vec{P}_{i}) = \int d\vec{p}_{M'}
d\vec{p}_{B'} d\vec{p}_{M} d\vec{p}_{B} d\vec{P} \phi_{M'}^{\ast}(\vec{p}_{M'}) \times \nonumber \\
&
\times  \phi_{B'}^{\ast}(\vec{p}_{B'})
{\cal H}(\vec{P}',\vec{P}) P_{mn} \left[\phi_M(\vec{p}_{M}) \phi_B(\vec{p}_{B}) \delta^{(3)}(\vec{P}-\vec{P}_{i}) \right] \nonumber\,,\\
&
{}^{\rm RGM}N_{E}(\vec{P}',\vec{P}_{i}) = \int d\vec{p}_{M'}
d\vec{p}_{B'} d\vec{p}_{M} d\vec{p}_{B} d\vec{P} \phi_{M'}^{\ast}(\vec{p}_{M'}) \times \nonumber \\
&
\times  \phi_{B'}^{\ast}(\vec{p}_{B'})
P_{mn} \left[\phi_M(\vec{p}_{M}) \phi_B(\vec{p}_{B}) \delta^{(3)}(\vec{P}-\vec{P}_{i}) \right] \,,
\end{align}
where ${\cal H}$ is the Hamiltonian at quark level and $P_{mn}$ is the exchange operators $P_{12}$ or $P_{13}$.

If we want to connect different channels, such as $\bar D^{(*)}\Xi_c$ and $J/\psi\Lambda$ channels, ${}^{\rm RGM}N_{E}$ does not affect and ${}^{\rm RGM}H_{E}$ can be reduced to an exchange potential,

\begin{align}\label{eq:exchangeV}
&
{}^{\rm RGM}V_{E}(\vec{P}',\vec{P}_{i}) = \sum_{i\in M, j\in B} \int d\vec{p}_{M'}
d\vec{p}_{B'} d\vec{p}_{M} d\vec{p}_{B} d\vec{P} \phi_{M}^{\ast}(\vec{p}_{M'}) \times \nonumber \\
&
\times  \phi_{B}^{\ast}(\vec{p}_{B'})
V_{ij}(\vec{P}',\vec{P}) P_{mn} \left[\phi_{M'}(\vec{p}_{M}) \phi_{B'}(\vec{p}_{B}) \delta^{(3)}(\vec{P}-\vec{P}_{i}) \right] \,.
\end{align}

In order to describe the meson $q\bar q$ and baryon $qqq$ bound states, we solve the
Schr\"odinger equation using the Gaussian Expansion Method~\cite{Hiyama:2003cu}. In this method, the radial wave function of the cluster is expanded in terms of Gaussian functions whose parameters are in geometrical progression.
For the radial meson wave function we can write

\begin{align}
 \phi_M^{\ell m}(\vec p) = \sum_{n}^{n_{\rm max}} C^{(M)}_{n} \phi_{n,\ell m}(\vec p)
\end{align}
where $\ell$ is the total angular momentum, $m$ its projection and $C^{(M)}_n$ the coefficients of the base expansion.
For the baryon, the radial wave function is similar, but we have two momenta for the $\rho$ and $\lambda$ modes,

\begin{align}
 \phi_B^{\ell m}(\vec p_\lambda,\vec p_\rho) = \sum_{n_\lambda,n_\rho}^{n_{\rm max}} C^{(B)}_{n_\lambda,n_\rho} \left[\phi_{n_\lambda,\ell_\lambda}(\vec p_\lambda)\phi_{n_\rho,\ell_\rho}(\vec p_\rho)\right]_{\ell m}
\end{align}
where $\ell=\ell_\lambda\oplus\ell_\rho$. In both cases, the $\phi$ functions are defined as

\begin{eqnarray*}
 \phi_{n,\ell m}(\vec p)&=& N_{n\ell}\, p^\ell\, e^{-\frac{1}{4\eta_n} p^2} Y_{\ell m}(\hat p).
\end{eqnarray*}
with $N_{n\ell}$ the normalization of the Gaussian wave functions such that $\langle\phi_{n\ell}|\phi_{n\ell}\rangle=1$.

The coefficients $C_n$ and the eigenenergies of the meson (baryon) are determined from the following set of equations,

\begin{align}
 \sum_{n=1}^{n_{\rm max}} \left[ (T_{n'n}-E N_{n'n})C_n +
 \sum_\alpha V_{n'n} C_n\right]= 0 
\end{align}
with $T_{n'n}$ and $N_{n'n}$ the kinetic and normalization operators, which are diagonal, and $V_{n'n}$ the underlying $qq$ interaction from the constituent quark model detailed above.

The masses and properties of meson-baryon molecular candidates are obtained from a coupled-channels calculation by searching the poles of the $S$-matrix,
calculated by means of the $T$-matrix from the Lippmann-Schwinger equation

\begin{align} \label{ec:Tonshell}
T_\beta^{\beta'}(z;p',p)&=V_\beta^{\beta'}(p',p)+\sum_{\beta''}\int dp''p''^2
V_{\beta''}^{\beta'}(p',p'')\nonumber\\
&\frac{1}{z-E_{\beta''}(p'')}T_{\beta}^{\beta''}(z;p'',p)
\end{align}
where $\beta$ represents the set of quantum numbers necessary to determine a
partial wave in the meson-baryon state, $V_{\beta}^{\beta'}(p',p)$ is the RGM potential
 and  $E_{\beta''}(p'')$ is the energy for the momentum $p''$
referred to the lower threshold.
Eq.~\eqref{ec:Tonshell} is solved using a generalized version of the matrix inversion
method~\cite{machleidt1993one} including channels with different thresholds, whereas poles are calculated by means of the Broyden method~\cite{broyden1965class}.

\section{Results} \label{sec:results}

\subsection{$P_{\psi s}^\Lambda$ sector}

The $P_{\psi s}^\Lambda(4459)^0$ was first seen in the $J/\psi\Lambda$ invariant mass spectrum~\cite{LHCb:2020jpq}. It is an isospin-0 resonance whose mass is between the $\bar D\Xi_c'$ and $\bar D^*\Xi_c$. Thus, it is reasonable to
analyze its assignment as a $\bar D^{(*)}\Xi_c^{(\prime)}$ molecule.
The most interesting sectors, which will be explored in this work, are $J^P=\frac{1}{2}^-$, $\frac{3}{2}^-$ and $\frac{5}{2}^-$, where the $\bar D^{(*)}$ and the $\Xi_c^{(\prime)(*)}$ are in a relative S-wave.
States with positive parity are not favoured as it forces the molecule to be in a relative P-wave.

\begin{table}
\begin{center}
\begin{tabular}{cc|cc|cc|cc}
\hline\hline
\multicolumn{2}{c|}{$P_{\psi s}^\Lambda$} & \multicolumn{2}{c|}{$P_{\psi s}^\Sigma$} & \multicolumn{2}{c|}{$P_{\Upsilon s}^\Lambda$} & \multicolumn{2}{c}{$P_{\Upsilon s}^\Sigma$} \\
Channel   & Mass & Channel   & Mass & Channel   & Mass & Channel   & Mass   \\
\hline
$\eta_c \Lambda$ &   $4099.1$  &  $\eta_c\Sigma$  &  $4176.5$  &  $\eta_b \Lambda$ &   $10514.7$  &  $\eta_b\Sigma$  &  $10592.1$ \\
$J/\psi \Lambda$ &   $4212.6$  &  $J/\psi\Sigma$  &  $4290.0$  &  $\Upsilon \Lambda$ &   $10576.0$  &  $\Upsilon\Sigma$  &  $10653.4$ \\
$\bar D_s\Lambda_c$  &  $4254.8$  &   $\bar D_s\Sigma_c$  &  $4421.9$   & $\bar B_s\Lambda_b$  &  $10986.5$  &   $\bar B_s\Sigma_b$  &  $11180.3$ \\
$\bar D_s^* \Lambda_c$ &   $4398.7$  &  $\bar D_s^*\Sigma_c$  &  $4565.7$  &  $\bar B_s^* \Lambda_b$ &   $11035.0$  &  $\bar B_s^*\Sigma_b$  &  $11228.8$ \\
$\bar D \Xi_c$ &   $4336.6$  &  $\bar D \Xi_c$ &   $4336.6$  &  $\bar B \Xi_b$ &   $11072.7$  &  $\bar B \Xi_b$ &   $11072.7$  \\
$\bar D \Xi'_c$ &  $4445.3$  &  $\bar D \Xi'_c$ &  $4445.3$  &  $\bar B \Xi'_b$ &  $11214.5$  &  $\bar B \Xi'_b$ &  $11214.5$ \\
$\bar D\Xi^*_c$ &  $4513.2$  &  $\bar D\Xi^*_c$ &  $4513.2$  &  $\bar B\Xi^*_b$ &  $11233.3$  &  $\bar B\Xi^*_b$ &  $11233.3$ \\
$\bar D^*\Xi_c$ &   $4477.9$  &  $\bar D^*\Xi_c$ &   $4477.9$  &  $\bar B^*\Xi_b$ &   $11117.9$  &  $\bar B^*\Xi_b$ &   $11117.9$  \\
$\bar D^*\Xi'_c$ & $4586.6$  &  $\bar D^*\Xi'_c$ & $4586.6$   &  $\bar B^*\Xi'_b$ & $11259.7$  &  $\bar B^*\Xi'_b$ & $11259.7$ \\
$\bar D^*\Xi^*_c$ &  $4654.5$  &  $\bar D^*\Xi^*_c$ &  $4654.5$  &  $\bar B^*\Xi^*_b$ &  $11278.5$  &  $\bar B^*\Xi^*_b$ &  $11278.5$ \\
\hline\hline
 \end{tabular}
\end{center}\caption{\label{tab:thres} Threshold masses (in MeV) of the meson-baryon channels considered in this work for the  isospin-0 $P_{\psi s}^\Lambda$ (minimum quark content $\bar cc snn$), $P_{\Upsilon s}^\Lambda$ ($\bar bb snn$) and isospin-1 $P_{\psi s}^\Sigma$ ($\bar ccsnn$) and $P_{\Upsilon s}^\Sigma$ ($\bar bbsnn$) states. }
\end{table}

Hence, first we perform a coupled-channels calculation of $J^P=\frac{1}{2}^-$, $\frac{3}{2}^-$ and $\frac{5}{2}^-$ sectors including the channels detailed in Tab.~\ref{tab:thres}. The $J/\psi\Lambda$ and $\eta_c\Lambda$ channels have a small influence on the pole formation, but they contribute to the decay channels.

 \begin{table}
  \begin{tabular}{cl|cc}\hline\hline
   $J^P$ & Assignment & Mass & Width  \\ \hline
   \multirow{6}{*}{$\frac{1}{2}^-$} &${\bf P_{\psi s}^\Lambda(4338)}$ & $4341.0$ & $14.0$   \\
    & ${\bf P_{\psi s}^\Lambda(4459)}$ & $4465.1$ & $24.1$  \\
   &$P_{\psi s}^\Lambda(4382)$ & $4381.7$ & $76.7$  \\
   &$P_{\psi s}^\Lambda(4443)$ & $4443.6$ & $0$  \\
   &$P_{\psi s}^\Lambda(4580)$ & $4581.0$ & $7.4$  \\
   &$P_{\psi s}^\Lambda(4647)$ & $4647.5$ & $2.7$  \\ \hline
   $\frac{3}{2}^-$ &$P_{\psi s}^\Lambda(4655)$ & $4655.3$ & $9.5$   \\
   \hline\hline
   \end{tabular}
  \caption{\label{tab:Pcs0} Masses and widths (in MeV) of the $P_{\psi s}^\Lambda$ states found in this work. The resonances assigned to experimental states are shown in bold, the rest of states are theoretical predictions not yet detected.}
 \end{table}

The states found in this calculation are shown in Tables~\ref{tab:Pcs0} (masses and widths) and~\ref{tab:Pcs1} (probabilities and partial widths). We find two resonances in the $J^P=\frac{1}{2}^-$ sector, with masses and widths compatible with the experimental $P_{\psi s}^\Lambda(4459)^0$ and $P_{\psi s}^\Lambda(4338)^0$.
The theoretical $P_{\psi s}^\Lambda(4459)^0$ resonance can be interpreted as a $\bar D^*\Xi_c$ molecule ($59.5\%$) with a large admixture of $\bar D_s^*\Lambda_c$ ($34.7\%$), which is its main decay channel. The $P_{\psi s}^\Lambda(4338)^0$ candidate, on the contrary, is mainly a $\bar D_s\Lambda_c$ molecule ($45\%$), due to the large coupling to this threshold, with a large admixture of $\bar D_s^*\Lambda_c$ ($28\%$). This large admixture to the $\bar D_s\Lambda_c$ and $\bar D_s^*\Lambda_c$ for the $P_{\psi s}^\Lambda(4459)^0$ and $P_{\psi s}^\Lambda(4338)^0$ was also predicted at Ref.~\cite{Xiao:2019gjd}.  The predicted mass of the $P_{\psi s}^\Lambda(4338)^0$ is in good agreement with the experimental value, just few MeV above it, due mainly to the inclusion of the $\bar D_s\Lambda_c$ threshold. 

  \begin{table*}
  \begin{tabular*}{\linewidth}{l @{\extracolsep{\fill}}|cccccccccc|cccccccc}\hline\hline
            &  \multicolumn{10}{c|}{Probabilities [\%]} & \multicolumn{8}{c}{Partial widths [MeV]} \\
   Assignment &  $ {\eta_c\Lambda}$ &  $ {J/\psi\Lambda}$ &  $ {\bar D_s\Lambda_c}$&  $ {\bar D\Xi_c}$ & $ {\bar D_s^*\Lambda_c}$& $ {\bar D\Xi_c'}$& $ {\bar D^*\Xi_c}$ & $ {\bar D\Xi_c^*}$&  $ {\bar D^*\Xi_c'}$ & $ {\bar D^*\Xi_c^*}$ & $ {\eta_c\Lambda}$& $ {J/\psi\Lambda}$& $ {\bar D_s\Lambda_c}$& $ {\bar D\Xi_c}$ & $ {\bar D_s^*\Lambda_c}$& $ {\bar D\Xi_c'}$ & $ {\bar D\Xi_c^*}$& $ {\bar D^*\Xi_c'}$\\ \hline
   ${\bf P_{\psi s}^\Lambda(4338)}$ & $7.3$ & $4.4$ & $45.0$ & $6.3$ & $28.0$ & $0$ & $9.0$ & $0$& $0$ & $0$ &   $1.2$ & $0.6$& $11.0$    & $1.1$    & $0$  & $0$ & $0$   & $0$ \\
   ${\bf P_{\psi s}^\Lambda(4459)}$ & $0.9$& $0.2$ & $1.3$ & $3.4$ & $34.7$ & $0$ & $59.5$& $0$ & $0$ & $0$ & $0.7$&  $1.6$ & $3.4$ & $1.9$ & $16.6$ & $0$ &  $0$  & $0$\\
   $P_{\psi s}^\Lambda(4382)$ & $3.5$ & $2.2$& $10.7$& $5.6$ & $55.8$ & $0$ & $22.2$ & $0$ & $0$ & $0$ & $2.1$&  $30.8$ & $43.2$ & $0.7$ & $0$& $0$ & $0$ & $0$\\
  $P_{\psi s}^\Lambda(4443)$ & $0$ & $0$ & $0$ & $0$ & $0$ & $99.5$ & $0$ & $0$& $0.4$ & $0.1$ & $0.1$&  $0.9$ & $0$ & $0$ & $0$ & $0$  & $0$ & $0$\\
   $P_{\psi s}^\Lambda(4580)$ & $0.2$ & $0.6$ & $0$ & $0$ & $0$ & $20.6$ & $0$ & $0$ & $78.3$ & $0.3$ & $0.1$&  $0.1$ & $0$ & $0$ & $0$ & $7.2$ & $0$& $0$\\
   $P_{\psi s}^\Lambda(4647)$ & $0.0$ & $0.0$ & $0$ & $0$ & $0$ & $18.0$ & $0$ & $0$ & $9.0$ & $73.0$ & $0.004$&  $0.002$ & $0$ & $0$ & $0$ & $2.0$  & $0$& $0.6$ \\
   $P_{\psi s}^\Lambda(4655)$ & $0$ & $0.0$ & $0$ & $0$ & $0$ & $4.8$ & $0$ & $25.2$ & $22.3$ & $47.8$ & $0.01$&  $0$ & $0$ & $0$ & $0$ & $0.3$  & $6.1$ & $3.1$ \\
   \hline\hline
   \end{tabular*}
  \caption{\label{tab:Pcs1} Probabilities and partial widths of open-charmed channels in the $P_{\psi s}^\Lambda$ states found in this work.}
 \end{table*}

The partial widths are also shown in Table~\ref{tab:Pcs1}. The $P_{\psi s}^\Lambda(4338)$ decays mainly to $\bar D_s\Lambda$, with also significant contributions to $\bar D\Xi_c$, $J/\psi\Lambda$ and $\eta_c\Lambda$. The width is, though, a bit larger than the experimental one, which could be explained by the large coupling to the $\bar D_s\Lambda_c$ channel.  
The $P_{\psi s}^\Lambda(4459)$ has large partial widths to the  $\bar D_s\Lambda$, $\bar D\Xi_c$ and $\bar D_s^*\Lambda_c$, which is its main decay channel. The $J/\psi\Lambda$ partial width is smaller, but relevant as it is the discovery channel. We want to remark that, as we employ a phenomenological model, systematic uncertainties cannot be evaluated, which should also be taken into account when comparing to experimental values.

Apart from the two experimentally confirmed states, $P_{\psi s}^\Lambda(4338)$ and $P_{\psi s}^\Lambda(4459)$, we find four additional $J^P=\frac{1}{2}^-$ and one $J^P=\frac{3}{2}^-$ $P_{\psi s}^\Lambda$.  The $J^P=\frac{3}{2}^-$ state, denoted as $P_{\psi s}^\Lambda(4655)$, is a relatively narrow resonance around the $\bar D^*\Xi_c^*$ thresholds. It decays mostly to $\bar D\Xi_c^*$ and $\bar D^*\Xi_c^\prime$. The $P_{\psi s}^\Lambda(4382)$ is a wide $\frac{1}{2}^-$ resonance just below the $\bar D_s^*\Lambda_c$ ($\sim\!\!56\%$), whose main decay channels are $J/\psi\Lambda$ and $\bar D_s\Lambda_c$.
The so-called $P_{\psi s}^\Lambda(4443)$, $P_{\psi s}^\Lambda(4580)$ and $P_{\psi s}^\Lambda(4647)$ are three $J^P=\frac{1}{2}^-$ molecules below the $\bar D\Xi_c^\prime$, $\bar D^*\Xi_c^\prime$ and $\bar D^*\Xi_c^*$ thresholds, respectively. They are relatively narrow, with widths below $10$ MeV, which decay to the lower $\bar D\Xi_c^\prime$ and $\bar D^*\Xi_c^\prime$ channels. It is interesting to mention that, in our model, such resonances cannot decay to channels with a $\Lambda_c$ or $\Xi_c$ due to the different flavor structure of the baryon wave function. For such ground-state baryons, the flavor wave function of the light diquark is in an antisymmetric state, whereas for $\Xi_c^\prime$ and $\Xi_c^*$, the diquark is in a symmetric state, and the interaction between these structures is not allowed in our model.

 \subsection{$P_{\psi s}^\Sigma$ sector}

 \begin{table}
  \begin{tabular}{cl|cc|ccccc}\hline\hline
   $J^P$ & Assignment & Mass & Width & $\Gamma_{J/\psi\Sigma}$  &  $\Gamma_{\bar D_s\Sigma_c}$ & $\Gamma_{\bar D\Xi_c'}$  & $\Gamma_{\bar D\Xi_c^*}$ \\ \hline
   $\frac{3}{2}^-$ & $P_{\psi s}^\Sigma(4547)$ & $4547.3$ & $26.72$ &   $0.1$      & $0.2$  & $0$     & $26.3$\\ 
   $\frac{5}{2}^-$ & $P_{\psi s}^\Sigma(4456)$ & $4456.8$ & $74.57$ & $0.01$     & $47.9$ & $26.6$    & $0$ \\
   \hline\hline
   \end{tabular}
  \caption{\label{tab:Pcs0I1} Masses, widths and partial widths (in MeV) of the $P_{\psi s}^\Sigma$ states found in this work. }
 \end{table}

 Additionally, we have explored the isospin $1$ sector of the $\bar ccsnn$ system, i.e. the $P_{\psi s}^\Sigma$ pentaquark states, with $J^P=\frac{1}{2}^-$, $\frac{3}{2}^-$ and $\frac{5}{2}^-$. The thresholds included in the coupled-channels calculation are mostly the same as for the isospin $0$ calculation (see Table~\ref{tab:thres}). The pentaquark candidates found in this sector are shown in Tables~\ref{tab:Pcs0I1} (masses, widths and non-zero partial widths). Only two candidates are found just below the $\bar D^*\Xi_c^*$ threshold, a resonance in the $\frac{3}{2}^-$ channel close to $\bar D_s^*\Sigma_c$ threshold, with $\sim\!27$ MeV width, and a wider resonance in $\frac{5}{2}^-$, above the $\bar D\Xi_c^\prime$ thresholds. The partial widths show that $J/\psi\Sigma$ is a viable detection channels for the $\frac{3}{2}^-$ molecule, but open-charmed channels are more convenient to detect the $\frac{5}{2}^-$ resonance. In this case, channels with $\Xi_c$ are disconnected from those containing a $\Sigma_c$, $\Xi_c^\prime$ or $\Xi_c^*$, due to the flavor symmetry of the light diquark in these baryons.

 \subsection{Hidden-bottom $P_{\Upsilon s}^\Lambda$ and $P_{\Upsilon s}^\Sigma$ sector}

 In this section we extend our analysis to the bottom sector, studying the hidden-bottom pentaquarks $P_{\Upsilon s}^\Lambda$ and $P_{\Upsilon s}^\Sigma$ states, with minimum quark content $\bar b b snn$ and isospin $0$ and $1$, respectively. Our model allows to analyze such structures with no further tuning of the parameters. The reduction of the kinetic energy of the baryon-meson system due to the larger bottom quark mass favors the formation of new molecules, thus we find more candidates than for the charm sector. 
 
 We, then, explore the $\frac{1}{2}^-$, $\frac{3}{2}^-$ and $\frac{5}{2}^-$ sectors, including the channels of Table~\ref{tab:thres}, which are the bottom analogs of the coupled-channels calculation done for the $P_{\psi s}^\Lambda$ and $P_{\psi s}^\Sigma$ sectors.  We obtain twelve $P_{\Upsilon s}^\Lambda$ candidates and eleven $P_{\Upsilon s}^\Sigma$ ones. Results are shown in Table~\ref{tab:Pbs0}, detailing the sector, mass, total width and main decay channel. For most of them, $\Upsilon(1S)\Lambda$ is a good detection channel, in analogy with the $J/\psi\Lambda$ channel where the $P_{\psi s}^\Lambda(4459)$ and $P_{\psi s}^\Lambda(4338)$ pentaquarks were first detected.

 \begin{table}
  \begin{tabular}{l|cccl}\hline\hline
&$J^P$ & Mass & Width & Main Decay Channel \\\hline
\multirow{12}{*}{$P_{\Upsilon s}^\Lambda$} & \multirow{6}{*}{$\frac{1}{2}^-$} & 10671.8 & 89.2 & $\Upsilon(1S)\Lambda$ (87.4) \\
& & 10994.8 & 5.2 & $B_s\Lambda_b$ (3.6) \\
& & 11208.0 & 44.5 & $B^*\Xi_c$ (33.1) \\
& & 11238.7 & 39.7 & $B^*\Xi_b$ (19.2) \\
& & 11042.6 & 59.0 & $B_s^*\Lambda_b$ (48.9) \\
& & 11098.5 & 83.6 & $B_s\Lambda_b$ (64.6) \\ \cline{2-5}
&\multirow{2}{*}{$\frac{3}{2}^-$} & 11090.4 & 66.3 & $B\Xi_b$ (56.8) \\
& & 11043.0 & 58.0 & $B_s^*\Lambda_b$ (56.4) \\
\cline{2-5}
&\multirow{4}{*}{$\frac{5}{2}^-$} & 10995.9 & 57.7 & $B_s\Lambda_b$ (54.9) \\
& & 11043.1 & 54.8 & $B_s^*\Lambda_b$ (45.0) \\
& & 11098.6 & 47.1 & $B\Xi_b$ (30.3) \\
& & 11141.7 & 49.1 & $B^*\Xi_b$ (29.8) \\
\hline       
\multirow{11}{*}{$P_{\Upsilon s}^\Sigma$} & \multirow{3}{*}{$\frac{1}{2}^-$} & 11074.8 & 52.0 & $B\Xi_b$ (52.0) \\
& & 11193.9 & 34.4 &  $B_s\Sigma_b$ (33.6) \\
& & 11132.4 & 66.2 & $B^*\Xi_b$ (66.1) \\
\cline{2-5}
&\multirow{3}{*}{$\frac{3}{2}^-$} & 11188.1 & 59.1 & $B_s\Sigma_b$ (58.8) \\
& & 11251.2 & 46.6 &  $B\Xi_b^*$ (23.9) \\
& & 11232.9 & 7.9 & $B_s^*\Sigma_b$ (5.6) \\
\cline{2-5}
&\multirow{5}{*}{$\frac{5}{2}^-$} & 11079.9 & 54.4 & $B\Xi_b$ (54.4) \\
& & 11125.2 & 52.9 & $B^*\Xi_b$ (52.9) \\
& & 11193.2 & 41.5 & $B_s\Sigma_b$ (41.4) \\
& & 11243.5 & 40.7 & $B_s^*\Sigma_b$ (39.9) \\
& & 11272.5 & 30.5 & $B_s^*\Sigma_b$ (17.4) \\
  \hline \hline
  \end{tabular}
  \caption{\label{tab:Pbs0} Masses, widths and main decay channel, with partial width in parenthesis (in MeV), of the $P_{\Upsilon s}^\Lambda$ and $P_{\Upsilon s}^\Sigma$ states found in this work. }
 \end{table}

\subsection{$P_{\psi ss}^N$ and $P_{\Upsilon ss}^N$ sectors}

The $P_{\psi ss}^N$ ($P_{\Upsilon ss}^N$) are pentaquark-like states with $\bar cc ssn$ ($\bar bbssn$) minimum quark content and $I=\frac{1}{2}$. Up to now, no
states of this kind have been experimentally discovered. Yet, the similarity with the $P_{\psi s}^\Lambda$ states
make them an interesting system that have been explored in the recent literature~\cite{Azizi:2021pbh,Ferretti:2020ewe,Wang:2020bjt, Wang:2022neq,Valera:2022elt}, pointing to a rich spectroscopy. Our constituent quark model allows to make predictions of such states
with no parameter tuning. The relevant thresholds for this system are given in Table~\ref{tab:thres2}.

In this sector, channels are mostly disconnected, as no $\pi$ exchange is allowed between them and other diagrams such as annihilation through a gluon are negligible. Then, the dynamics is mostly governed by $\sigma$ exchanges, which gives diagonal contributions. So, in this case, there is no gain in performing a coupled-channels calculation, and an analysis of individual thresholds can be considered as a good approach.

\begin{table}
\begin{center}
\begin{tabular}{cc|cc}
\hline\hline
\multicolumn{2}{c|}{$P_{\psi ss}^N$} & \multicolumn{2}{|c}{$P_{\Upsilon ss}^N$} \\
Channel &   Mass [MeV] & Channel &   Mass [MeV] \\
\hline
$\bar D_s \Xi_c$&  $4437.72$ & $\bar B_s \Xi_b$&  $11160.08$  \\
$\bar D_s \Xi'_c$&  $4546.45$ & $\bar B_s \Xi'_b$&  $11301.90$  \\
$\bar D_s^*\Xi_c$& $4581.57$ & $\bar B_s^*\Xi_b$& $11208.60$ \\
$\bar D_s\Xi_c^*$& $4614.27$ & $\bar B_s\Xi_b^*$& $11320.70$ \\
$\bar D_s^*\Xi'_c$& $4690.30$ & $\bar B_s^*\Xi'_b$& $11350.42$ \\
$\bar D_s^*\Xi_c^*$& $4758.12$ & $\bar B_s^*\Xi_b^*$& $11369.21$ \\
\hline\hline
 \end{tabular}
\end{center}\caption{\label{tab:thres2} Masses of the meson-baryon channels considered in this work for the $P_{\psi ss}^N$ (minimum quark content $\bar cc ssn$) and $P_{\Upsilon ss}^N$ (minimum quark content $\bar bb ssn$).}
\end{table}

Results are shown in Table~\ref{tab:Pcss}. In the hidden-charm sector we find eight $P_{\psi ss}^N$ meson-baryon molecules: three $\frac{1}{2}^-$, four $\frac{3}{2}^-$ and one $\frac{5}{2}^-$.
Most of the candidates are potentially detectable in the $J/\psi\Xi$ or $\eta_c\Xi$ channels, the analogs of the $J/\psi\Lambda$ and $\eta_c\Lambda$ in the $P_{\psi ss}^N$ sector. For the hidden-bottom sector we find nine $P_{\Upsilon ss}^N$ molecules, with larger binding energies. The main decay channels for these candidates are mostly $\Upsilon\Xi^{(*)}$ and $\eta_b\Xi^{(*)}$.

 \begin{table}
  \begin{tabular}{l|cccl}\hline\hline
& $J^P$ &  Channel & Mass [MeV] &  Main Decay channels\\ \hline
\multirow{8}{*}{$P_{\psi ss}^N$} &  \multirow{3}{*}{$\frac{1}{2}^-$} & $\bar D_s\Xi_c$ & $4436.8$ &  $J/\psi\Xi$ ($9.9$), $\eta_c\Xi$ ($1.4$)\\
&   & $\bar D_s\Xi_c'$ & $4544.0$ &  $J/\psi\Xi$ ($1.0$), $\eta_c\Xi$ ($0.7$) \\
&   & $\bar D_s^*\Xi_c$ & $4580.8$ &  $J/\psi\Xi$ ($0.9$),  $\eta_c\Xi$ ($1.0$) \\ \cline{2-5}
&   \multirow{4}{*}{$\frac{3}{2}^-$} & $\bar D_s^*\Xi_c$ & $4581.1$ &  $J/\psi\Xi$ ($1.2$) \\
&    & $\bar D_s\Xi_c^*$ & $4613.0$ &  $J/\psi\Xi$ ($0.8$), $\eta_c\Xi^*$ ($1.3$) \\
&    & $\bar D_s^*\Xi_c'$ & $4684.9$ &  $\bar D^*_s\Xi_c^*$ ($3.1$), $J/\psi\Xi^*$ ($7.7$)\\
&    & $\bar D_s^*\Xi_c^*$ & $4758.1$ &  $\eta_c\Xi^*$ ($0.03$) \\ \cline{2-5}
&  $\frac{5}{2}^-$ & $\bar D_s^*\Xi_c^*$ & $4751.7$ &  $J/\psi\Xi^*$ ($4.6$)  \\
  \hline 
\multirow{9}{*}{$P_{\Upsilon ss}^N$}  &  \multirow{4}{*}{$\frac{1}{2}^-$} & $B_s\Xi_b$ & $11142.6$ & $\Upsilon\Xi$ ($1.3$), $\eta_b\Xi$ ($0.2$)\\
&   & $B_s\Xi_b'$ & $11281.4$ &  $\Upsilon\Xi^*$ ($2.0$) \\
&   & $B_s^*\Xi_b'$ & $11348.5$ &  $B_s\Xi_b'$ ($22.4$) \\
&   & $B_s^*\Xi_b$ & $11191.9$ &  $\Upsilon\Xi$ ($0.2$),  $\eta_b\Xi$ ($0.4$) \\ \cline{2-5}
&   \multirow{4}{*}{$\frac{3}{2}^-$} & $B_s^*\Xi_b$ & $11194.2$ &  $\Upsilon\Xi$ ($0.3$) \\
&    & $B_s\Xi_b^*$ & $11304.0$ &  $\Upsilon\Xi*$ ($0.5$), $\eta_b\Xi^*$ ($0.1$) \\
&    & $B_s^*\Xi_b'$ & $11321.8$ &  $B^*_s\Xi_b^*$ ($5.7$), $\Upsilon\Xi^*$ ($0.6$)\\
&    & $B_s^*\Xi_b^*$ & $11361.9$ &  $B_s\Xi_c^*$ ($0.4$)  \\ \cline{2-5}
&  $\frac{5}{2}^-$ & $B_s^*\Xi_b^*$ & $11337.6$ &  $\Upsilon\Xi^*$ ($0.5$)  \\
  \hline \hline
  \end{tabular}
  \caption{\label{tab:Pcss} Predicted $P_{\psi ss}^N$ and $P_{\Upsilon ss}^N$ meson-baryon molecules. The main decay channels of each candidate is shown, with the partial width (in MeV) in parenthesis. }
 \end{table}

\section{Summary}\label{sec:summary}

In this work we have performed a coupled-channels calculation of the $P_{\psi s}^\Lambda$, $P_{\psi s}^\Sigma$ and $P_{\psi ss}^N$ hidden-charm pentaquark candidates, and their bottom partners $P_{\Upsilon s}^\Lambda$, $P_{\Upsilon s}^\Sigma$ and $P_{\Upsilon ss}^N$,  as molecular states in the framework of a constituent quark model that satisfactorily describes the $P_{\psi}^N$ states~\cite{Ortega:2016syt}.
All the states presented in this work are predictions of the model, as all the parameters are constrained from previous studies.

We find that the $P_{\psi s}^\Lambda(4338)$ and $P_{\psi s}^\Lambda(4459)$ experimental states can be described as $(I)J^P=(0)\frac{1}{2}^-$ baryon-meson molecules with minimum quark content $\bar ccsnn$. Along with this experimental states, further $\bar c c snn$ molecules are predicted: four additional $\frac{1}{2}^-$ and one $\frac{3}{2}^-$ $P_{\psi s}^\Lambda$ molecules. Their properties (mass, width, probabilities and partial widths) are given, which could be useful for their experimental detection. 
Additionally, we have explored the isospin-$1$ partners, the $P_{\psi s}^\Sigma$ pentaquarks, obtaining a narrow $\frac{3}{2}^-$ resonance and a wide $\frac{5}{2}^-$ one. For the bottom sector, $P_{\Upsilon s}^\Lambda$ and $P_{\Upsilon s}^\Sigma$, we obtain a rich spectroscopy, as a result of the reduction of the kinetic energy of the meson-baryon system due to the larger bottom quark mass. 

In the $P_{\psi ss}^N$ ($P_{\Upsilon ss}^N$) sector, that is, structures with isospin $\frac{1}{2}$ and minimum quark content $\bar ccssn$ ($\bar bbssn$), we find up to eight molecular candidates as $\bar D_s^{(*)}\Xi_c^{(*)(\prime)}$ molecules and nine molecular candidates as $B_s^{(*)}\Xi_b^{(*)(\prime)}$ states, which can be detected in future LHCb searches.

\begin{acknowledgments}
This work has been partially funded by Ministerio de Ciencia, Innovación y Universidades under Contract No.~PID2019-105439GB-C22/AEI/10.13039/501100011033,
and by the EU Horizon 2020 research and innovation program, STRONG-2020 project, under grant agreement No. 824093.
\end{acknowledgments}


\bibliographystyle{apsrev4-1}
\bibliography{Pcspaper}

\begin{thebibliography}{42}%
\makeatletter
\providecommand \@ifxundefined [1]{%
 \@ifx{#1\undefined}
}%
\providecommand \@ifnum [1]{%
 \ifnum #1\expandafter \@firstoftwo
 \else \expandafter \@secondoftwo
 \fi
}%
\providecommand \@ifx [1]{%
 \ifx #1\expandafter \@firstoftwo
 \else \expandafter \@secondoftwo
 \fi
}%
\providecommand \natexlab [1]{#1}%
\providecommand \enquote  [1]{``#1''}%
\providecommand \bibnamefont  [1]{#1}%
\providecommand \bibfnamefont [1]{#1}%
\providecommand \citenamefont [1]{#1}%
\providecommand \href@noop [0]{\@secondoftwo}%
\providecommand \href [0]{\begingroup \@sanitize@url \@href}%
\providecommand \@href[1]{\@@startlink{#1}\@@href}%
\providecommand \@@href[1]{\endgroup#1\@@endlink}%
\providecommand \@sanitize@url [0]{\catcode `\\12\catcode `\$12\catcode
  `\&12\catcode `\#12\catcode `\^12\catcode `\_12\catcode `\%12\relax}%
\providecommand \@@startlink[1]{}%
\providecommand \@@endlink[0]{}%
\providecommand \url  [0]{\begingroup\@sanitize@url \@url }%
\providecommand \@url [1]{\endgroup\@href {#1}{\urlprefix }}%
\providecommand \urlprefix  [0]{URL }%
\providecommand \Eprint [0]{\href }%
\providecommand \doibase [0]{http://dx.doi.org/}%
\providecommand \selectlanguage [0]{\@gobble}%
\providecommand \bibinfo  [0]{\@secondoftwo}%
\providecommand \bibfield  [0]{\@secondoftwo}%
\providecommand \translation [1]{[#1]}%
\providecommand \BibitemOpen [0]{}%
\providecommand \bibitemStop [0]{}%
\providecommand \bibitemNoStop [0]{.\EOS\space}%
\providecommand \EOS [0]{\spacefactor3000\relax}%
\providecommand \BibitemShut  [1]{\csname bibitem#1\endcsname}%
\let\auto@bib@innerbib\@empty
\bibitem [{\citenamefont {Aaij}\ \emph {et~al.}(2015)\citenamefont {Aaij} \emph
  {et~al.}}]{LHCb:2015yax}%
  \BibitemOpen
  \bibfield  {author} {\bibinfo {author} {\bibfnamefont {R.}~\bibnamefont
  {Aaij}} \emph {et~al.} (\bibinfo {collaboration} {LHCb}),\ }\href {\doibase
  10.1103/PhysRevLett.115.072001} {\bibfield  {journal} {\bibinfo  {journal}
  {Phys. Rev. Lett.}\ }\textbf {\bibinfo {volume} {115}},\ \bibinfo {pages}
  {072001} (\bibinfo {year} {2015})},\ \Eprint
  {http://arxiv.org/abs/1507.03414} {arXiv:1507.03414 [hep-ex]} \BibitemShut
  {NoStop}%
\bibitem [{\citenamefont {Aaij}\ \emph {et~al.}(2019)\citenamefont {Aaij} \emph
  {et~al.}}]{LHCb:2019kea}%
  \BibitemOpen
  \bibfield  {author} {\bibinfo {author} {\bibfnamefont {R.}~\bibnamefont
  {Aaij}} \emph {et~al.} (\bibinfo {collaboration} {LHCb}),\ }\href {\doibase
  10.1103/PhysRevLett.122.222001} {\bibfield  {journal} {\bibinfo  {journal}
  {Phys. Rev. Lett.}\ }\textbf {\bibinfo {volume} {122}},\ \bibinfo {pages}
  {222001} (\bibinfo {year} {2019})},\ \Eprint
  {http://arxiv.org/abs/1904.03947} {arXiv:1904.03947 [hep-ex]} \BibitemShut
  {NoStop}%
\bibitem [{\citenamefont {Wu}\ \emph {et~al.}(2010)\citenamefont {Wu},
  \citenamefont {Molina}, \citenamefont {Oset},\ and\ \citenamefont
  {Zou}}]{PhysRevLett.105.232001}%
  \BibitemOpen
  \bibfield  {author} {\bibinfo {author} {\bibfnamefont {J.-J.}\ \bibnamefont
  {Wu}}, \bibinfo {author} {\bibfnamefont {R.}~\bibnamefont {Molina}}, \bibinfo
  {author} {\bibfnamefont {E.}~\bibnamefont {Oset}}, \ and\ \bibinfo {author}
  {\bibfnamefont {B.~S.}\ \bibnamefont {Zou}},\ }\href {\doibase
  10.1103/PhysRevLett.105.232001} {\bibfield  {journal} {\bibinfo  {journal}
  {Phys. Rev. Lett.}\ }\textbf {\bibinfo {volume} {105}},\ \bibinfo {pages}
  {232001} (\bibinfo {year} {2010})}\BibitemShut {NoStop}%
\bibitem [{\citenamefont {Chen}\ \emph {et~al.}(2017)\citenamefont {Chen},
  \citenamefont {He},\ and\ \citenamefont {Liu}}]{Chen:2016ryt}%
  \BibitemOpen
  \bibfield  {author} {\bibinfo {author} {\bibfnamefont {R.}~\bibnamefont
  {Chen}}, \bibinfo {author} {\bibfnamefont {J.}~\bibnamefont {He}}, \ and\
  \bibinfo {author} {\bibfnamefont {X.}~\bibnamefont {Liu}},\ }\href {\doibase
  10.1088/1674-1137/41/10/103105} {\bibfield  {journal} {\bibinfo  {journal}
  {Chin. Phys. C}\ }\textbf {\bibinfo {volume} {41}},\ \bibinfo {pages}
  {103105} (\bibinfo {year} {2017})},\ \Eprint
  {http://arxiv.org/abs/1609.03235} {arXiv:1609.03235 [hep-ph]} \BibitemShut
  {NoStop}%
\bibitem [{\citenamefont {Santopinto}\ and\ \citenamefont
  {Giachino}(2017)}]{Santopinto:2016pkp}%
  \BibitemOpen
  \bibfield  {author} {\bibinfo {author} {\bibfnamefont {E.}~\bibnamefont
  {Santopinto}}\ and\ \bibinfo {author} {\bibfnamefont {A.}~\bibnamefont
  {Giachino}},\ }\href {\doibase 10.1103/PhysRevD.96.014014} {\bibfield
  {journal} {\bibinfo  {journal} {Phys. Rev. D}\ }\textbf {\bibinfo {volume}
  {96}},\ \bibinfo {pages} {014014} (\bibinfo {year} {2017})},\ \Eprint
  {http://arxiv.org/abs/1604.03769} {arXiv:1604.03769 [hep-ph]} \BibitemShut
  {NoStop}%
\bibitem [{\citenamefont {Xiao}\ \emph {et~al.}(2019)\citenamefont {Xiao},
  \citenamefont {Nieves},\ and\ \citenamefont {Oset}}]{Xiao:2019gjd}%
  \BibitemOpen
  \bibfield  {author} {\bibinfo {author} {\bibfnamefont {C.~W.}\ \bibnamefont
  {Xiao}}, \bibinfo {author} {\bibfnamefont {J.}~\bibnamefont {Nieves}}, \ and\
  \bibinfo {author} {\bibfnamefont {E.}~\bibnamefont {Oset}},\ }\href {\doibase
  10.1016/j.physletb.2019.135051} {\bibfield  {journal} {\bibinfo  {journal}
  {Phys. Lett. B}\ }\textbf {\bibinfo {volume} {799}},\ \bibinfo {pages}
  {135051} (\bibinfo {year} {2019})},\ \Eprint
  {http://arxiv.org/abs/1906.09010} {arXiv:1906.09010 [hep-ph]} \BibitemShut
  {NoStop}%
\bibitem [{\citenamefont {Aaij}\ \emph {et~al.}(2021)\citenamefont {Aaij} \emph
  {et~al.}}]{LHCb:2020jpq}%
  \BibitemOpen
  \bibfield  {author} {\bibinfo {author} {\bibfnamefont {R.}~\bibnamefont
  {Aaij}} \emph {et~al.} (\bibinfo {collaboration} {LHCb}),\ }\href {\doibase
  10.1016/j.scib.2021.02.030} {\bibfield  {journal} {\bibinfo  {journal} {Sci.
  Bull.}\ }\textbf {\bibinfo {volume} {66}},\ \bibinfo {pages} {1278} (\bibinfo
  {year} {2021})},\ \Eprint {http://arxiv.org/abs/2012.10380} {arXiv:2012.10380
  [hep-ex]} \BibitemShut {NoStop}%
\bibitem [{\citenamefont {Gershon}(2022)}]{Gershon:2022xnn}%
  \BibitemOpen
  \bibfield  {author} {\bibinfo {author} {\bibfnamefont {T.}~\bibnamefont
  {Gershon}} (\bibinfo {collaboration} {LHCb}),\ }\href@noop {} {\  (\bibinfo
  {year} {2022})},\ \Eprint {http://arxiv.org/abs/2206.15233} {arXiv:2206.15233
  [hep-ex]} \BibitemShut {NoStop}%
\bibitem [{\citenamefont {Chen}\ \emph {et~al.}(2021)\citenamefont {Chen},
  \citenamefont {Chen}, \citenamefont {Liu},\ and\ \citenamefont
  {Liu}}]{Chen:2020uif}%
  \BibitemOpen
  \bibfield  {author} {\bibinfo {author} {\bibfnamefont {H.-X.}\ \bibnamefont
  {Chen}}, \bibinfo {author} {\bibfnamefont {W.}~\bibnamefont {Chen}}, \bibinfo
  {author} {\bibfnamefont {X.}~\bibnamefont {Liu}}, \ and\ \bibinfo {author}
  {\bibfnamefont {X.-H.}\ \bibnamefont {Liu}},\ }\href {\doibase
  10.1140/epjc/s10052-021-09196-4} {\bibfield  {journal} {\bibinfo  {journal}
  {Eur. Phys. J. C}\ }\textbf {\bibinfo {volume} {81}},\ \bibinfo {pages} {409}
  (\bibinfo {year} {2021})},\ \Eprint {http://arxiv.org/abs/2011.01079}
  {arXiv:2011.01079 [hep-ph]} \BibitemShut {NoStop}%
\bibitem [{\citenamefont {Peng}\ \emph {et~al.}(2021)\citenamefont {Peng},
  \citenamefont {Yan}, \citenamefont {S\'anchez~S\'anchez},\ and\ \citenamefont
  {Valderrama}}]{Peng:2020hql}%
  \BibitemOpen
  \bibfield  {author} {\bibinfo {author} {\bibfnamefont {F.-Z.}\ \bibnamefont
  {Peng}}, \bibinfo {author} {\bibfnamefont {M.-J.}\ \bibnamefont {Yan}},
  \bibinfo {author} {\bibfnamefont {M.}~\bibnamefont {S\'anchez~S\'anchez}}, \
  and\ \bibinfo {author} {\bibfnamefont {M.~P.}\ \bibnamefont {Valderrama}},\
  }\href {\doibase 10.1140/epjc/s10052-021-09416-x} {\bibfield  {journal}
  {\bibinfo  {journal} {Eur. Phys. J. C}\ }\textbf {\bibinfo {volume} {81}},\
  \bibinfo {pages} {666} (\bibinfo {year} {2021})},\ \Eprint
  {http://arxiv.org/abs/2011.01915} {arXiv:2011.01915 [hep-ph]} \BibitemShut
  {NoStop}%
\bibitem [{\citenamefont {Liu}\ \emph {et~al.}(2021)\citenamefont {Liu},
  \citenamefont {Pan},\ and\ \citenamefont {Geng}}]{PhysRevD.103.034003}%
  \BibitemOpen
  \bibfield  {author} {\bibinfo {author} {\bibfnamefont {M.-Z.}\ \bibnamefont
  {Liu}}, \bibinfo {author} {\bibfnamefont {Y.-W.}\ \bibnamefont {Pan}}, \ and\
  \bibinfo {author} {\bibfnamefont {L.-S.}\ \bibnamefont {Geng}},\ }\href
  {\doibase 10.1103/PhysRevD.103.034003} {\bibfield  {journal} {\bibinfo
  {journal} {Phys. Rev. D}\ }\textbf {\bibinfo {volume} {103}},\ \bibinfo
  {pages} {034003} (\bibinfo {year} {2021})}\BibitemShut {NoStop}%
\bibitem [{\citenamefont {Chen}(2021)}]{Chen:2020kco}%
  \BibitemOpen
  \bibfield  {author} {\bibinfo {author} {\bibfnamefont {R.}~\bibnamefont
  {Chen}},\ }\href {\doibase 10.1103/PhysRevD.103.054007} {\bibfield  {journal}
  {\bibinfo  {journal} {Phys. Rev. D}\ }\textbf {\bibinfo {volume} {103}},\
  \bibinfo {pages} {054007} (\bibinfo {year} {2021})},\ \Eprint
  {http://arxiv.org/abs/2011.07214} {arXiv:2011.07214 [hep-ph]} \BibitemShut
  {NoStop}%
\bibitem [{\citenamefont {Karliner}\ and\ \citenamefont
  {Rosner}(2022)}]{Karliner:2022erb}%
  \BibitemOpen
  \bibfield  {author} {\bibinfo {author} {\bibfnamefont {M.}~\bibnamefont
  {Karliner}}\ and\ \bibinfo {author} {\bibfnamefont {J.~R.}\ \bibnamefont
  {Rosner}},\ }\href@noop {} {\  (\bibinfo {year} {2022})},\ \Eprint
  {http://arxiv.org/abs/2207.07581} {arXiv:2207.07581 [hep-ph]} \BibitemShut
  {NoStop}%
\bibitem [{\citenamefont {Hofmann}\ and\ \citenamefont
  {Lutz}(2005)}]{Hofmann:2005sw}%
  \BibitemOpen
  \bibfield  {author} {\bibinfo {author} {\bibfnamefont {J.}~\bibnamefont
  {Hofmann}}\ and\ \bibinfo {author} {\bibfnamefont {M.~F.~M.}\ \bibnamefont
  {Lutz}},\ }\href {\doibase 10.1016/j.nuclphysa.2005.08.022} {\bibfield
  {journal} {\bibinfo  {journal} {Nucl. Phys. A}\ }\textbf {\bibinfo {volume}
  {763}},\ \bibinfo {pages} {90} (\bibinfo {year} {2005})},\ \Eprint
  {http://arxiv.org/abs/hep-ph/0507071} {arXiv:hep-ph/0507071} \BibitemShut
  {NoStop}%
\bibitem [{\citenamefont {Anisovich}\ \emph {et~al.}(2015)\citenamefont
  {Anisovich}, \citenamefont {Matveev}, \citenamefont {Nyiri}, \citenamefont
  {Sarantsev},\ and\ \citenamefont {Semenova}}]{Anisovich:2015zqa}%
  \BibitemOpen
  \bibfield  {author} {\bibinfo {author} {\bibfnamefont {V.~V.}\ \bibnamefont
  {Anisovich}}, \bibinfo {author} {\bibfnamefont {M.~A.}\ \bibnamefont
  {Matveev}}, \bibinfo {author} {\bibfnamefont {J.}~\bibnamefont {Nyiri}},
  \bibinfo {author} {\bibfnamefont {A.~V.}\ \bibnamefont {Sarantsev}}, \ and\
  \bibinfo {author} {\bibfnamefont {A.~N.}\ \bibnamefont {Semenova}},\ }\href
  {\doibase 10.1142/S0217751X15501900} {\bibfield  {journal} {\bibinfo
  {journal} {Int. J. Mod. Phys. A}\ }\textbf {\bibinfo {volume} {30}},\
  \bibinfo {pages} {1550190} (\bibinfo {year} {2015})},\ \Eprint
  {http://arxiv.org/abs/1509.04898} {arXiv:1509.04898 [hep-ph]} \BibitemShut
  {NoStop}%
\bibitem [{\citenamefont {Feijoo}\ \emph {et~al.}(2016)\citenamefont {Feijoo},
  \citenamefont {Magas}, \citenamefont {Ramos},\ and\ \citenamefont
  {Oset}}]{Feijoo:2015kts}%
  \BibitemOpen
  \bibfield  {author} {\bibinfo {author} {\bibfnamefont {A.}~\bibnamefont
  {Feijoo}}, \bibinfo {author} {\bibfnamefont {V.~K.}\ \bibnamefont {Magas}},
  \bibinfo {author} {\bibfnamefont {A.}~\bibnamefont {Ramos}}, \ and\ \bibinfo
  {author} {\bibfnamefont {E.}~\bibnamefont {Oset}},\ }\href {\doibase
  10.1140/epjc/s10052-016-4302-7} {\bibfield  {journal} {\bibinfo  {journal}
  {Eur. Phys. J. C}\ }\textbf {\bibinfo {volume} {76}},\ \bibinfo {pages} {446}
  (\bibinfo {year} {2016})},\ \Eprint {http://arxiv.org/abs/1512.08152}
  {arXiv:1512.08152 [hep-ph]} \BibitemShut {NoStop}%
\bibitem [{\citenamefont {Lu}\ \emph {et~al.}(2016)\citenamefont {Lu},
  \citenamefont {Wang}, \citenamefont {Xie}, \citenamefont {Geng},\ and\
  \citenamefont {Oset}}]{Lu:2016roh}%
  \BibitemOpen
  \bibfield  {author} {\bibinfo {author} {\bibfnamefont {J.-X.}\ \bibnamefont
  {Lu}}, \bibinfo {author} {\bibfnamefont {E.}~\bibnamefont {Wang}}, \bibinfo
  {author} {\bibfnamefont {J.-J.}\ \bibnamefont {Xie}}, \bibinfo {author}
  {\bibfnamefont {L.-S.}\ \bibnamefont {Geng}}, \ and\ \bibinfo {author}
  {\bibfnamefont {E.}~\bibnamefont {Oset}},\ }\href {\doibase
  10.1103/PhysRevD.93.094009} {\bibfield  {journal} {\bibinfo  {journal} {Phys.
  Rev. D}\ }\textbf {\bibinfo {volume} {93}},\ \bibinfo {pages} {094009}
  (\bibinfo {year} {2016})},\ \Eprint {http://arxiv.org/abs/1601.00075}
  {arXiv:1601.00075 [hep-ph]} \BibitemShut {NoStop}%
\bibitem [{\citenamefont {Zhang}\ \emph {et~al.}(2020)\citenamefont {Zhang},
  \citenamefont {He},\ and\ \citenamefont {Ping}}]{Zhang:2020cdi}%
  \BibitemOpen
  \bibfield  {author} {\bibinfo {author} {\bibfnamefont {Q.}~\bibnamefont
  {Zhang}}, \bibinfo {author} {\bibfnamefont {B.-R.}\ \bibnamefont {He}}, \
  and\ \bibinfo {author} {\bibfnamefont {J.-L.}\ \bibnamefont {Ping}},\
  }\href@noop {} {\  (\bibinfo {year} {2020})},\ \Eprint
  {http://arxiv.org/abs/2006.01042} {arXiv:2006.01042 [hep-ph]} \BibitemShut
  {NoStop}%
\bibitem [{\citenamefont {Hu}\ and\ \citenamefont {Ping}(2022)}]{Hu:2021nvs}%
  \BibitemOpen
  \bibfield  {author} {\bibinfo {author} {\bibfnamefont {X.}~\bibnamefont
  {Hu}}\ and\ \bibinfo {author} {\bibfnamefont {J.}~\bibnamefont {Ping}},\
  }\href {\doibase 10.1140/epjc/s10052-022-10047-z} {\bibfield  {journal}
  {\bibinfo  {journal} {Eur. Phys. J. C}\ }\textbf {\bibinfo {volume} {82}},\
  \bibinfo {pages} {118} (\bibinfo {year} {2022})},\ \Eprint
  {http://arxiv.org/abs/2109.09972} {arXiv:2109.09972 [hep-ph]} \BibitemShut
  {NoStop}%
\bibitem [{\citenamefont {Xiao}\ \emph {et~al.}(2021)\citenamefont {Xiao},
  \citenamefont {Wu},\ and\ \citenamefont {Zou}}]{Xiao:2021rgp}%
  \BibitemOpen
  \bibfield  {author} {\bibinfo {author} {\bibfnamefont {C.~W.}\ \bibnamefont
  {Xiao}}, \bibinfo {author} {\bibfnamefont {J.~J.}\ \bibnamefont {Wu}}, \ and\
  \bibinfo {author} {\bibfnamefont {B.~S.}\ \bibnamefont {Zou}},\ }\href
  {\doibase 10.1103/PhysRevD.103.054016} {\bibfield  {journal} {\bibinfo
  {journal} {Phys. Rev. D}\ }\textbf {\bibinfo {volume} {103}},\ \bibinfo
  {pages} {054016} (\bibinfo {year} {2021})},\ \Eprint
  {http://arxiv.org/abs/2102.02607} {arXiv:2102.02607 [hep-ph]} \BibitemShut
  {NoStop}%
\bibitem [{\citenamefont {Zhu}\ \emph {et~al.}(2021)\citenamefont {Zhu},
  \citenamefont {Song},\ and\ \citenamefont {He}}]{Zhu:2021lhd}%
  \BibitemOpen
  \bibfield  {author} {\bibinfo {author} {\bibfnamefont {J.-T.}\ \bibnamefont
  {Zhu}}, \bibinfo {author} {\bibfnamefont {L.-Q.}\ \bibnamefont {Song}}, \
  and\ \bibinfo {author} {\bibfnamefont {J.}~\bibnamefont {He}},\ }\href
  {\doibase 10.1103/PhysRevD.103.074007} {\bibfield  {journal} {\bibinfo
  {journal} {Phys. Rev. D}\ }\textbf {\bibinfo {volume} {103}},\ \bibinfo
  {pages} {074007} (\bibinfo {year} {2021})},\ \Eprint
  {http://arxiv.org/abs/2101.12441} {arXiv:2101.12441 [hep-ph]} \BibitemShut
  {NoStop}%
\bibitem [{\citenamefont {Weng}\ \emph {et~al.}(2019)\citenamefont {Weng},
  \citenamefont {Chen}, \citenamefont {Deng},\ and\ \citenamefont
  {Zhu}}]{Weng:2019ynv}%
  \BibitemOpen
  \bibfield  {author} {\bibinfo {author} {\bibfnamefont {X.-Z.}\ \bibnamefont
  {Weng}}, \bibinfo {author} {\bibfnamefont {X.-L.}\ \bibnamefont {Chen}},
  \bibinfo {author} {\bibfnamefont {W.-Z.}\ \bibnamefont {Deng}}, \ and\
  \bibinfo {author} {\bibfnamefont {S.-L.}\ \bibnamefont {Zhu}},\ }\href
  {\doibase 10.1103/PhysRevD.100.016014} {\bibfield  {journal} {\bibinfo
  {journal} {Phys. Rev. D}\ }\textbf {\bibinfo {volume} {100}},\ \bibinfo
  {pages} {016014} (\bibinfo {year} {2019})},\ \Eprint
  {http://arxiv.org/abs/1904.09891} {arXiv:1904.09891 [hep-ph]} \BibitemShut
  {NoStop}%
\bibitem [{\citenamefont {Wang}\ and\ \citenamefont
  {Wang}(2022{\natexlab{a}})}]{Wang:2022gfb}%
  \BibitemOpen
  \bibfield  {author} {\bibinfo {author} {\bibfnamefont {X.-W.}\ \bibnamefont
  {Wang}}\ and\ \bibinfo {author} {\bibfnamefont {Z.-G.}\ \bibnamefont
  {Wang}},\ }\href@noop {} {\  (\bibinfo {year} {2022}{\natexlab{a}})},\
  \Eprint {http://arxiv.org/abs/2205.02530} {arXiv:2205.02530 [hep-ph]}
  \BibitemShut {NoStop}%
\bibitem [{\citenamefont {Chen}\ and\ \citenamefont
  {Sparado~Norella}(2022)}]{LHCbtalk}%
  \BibitemOpen
  \bibfield  {author} {\bibinfo {author} {\bibfnamefont {C.}~\bibnamefont
  {Chen}}\ and\ \bibinfo {author} {\bibfnamefont {E.}~\bibnamefont
  {Sparado~Norella}},\ }\href
  {https://indico.cern.ch/event/1176505/attachments/2475130/4248283/CERN%20seminar_LHCb.pdf}
  {\  (\bibinfo {year} {2022})},\ \Eprint
  {http://arxiv.org/abs/LHCb-PAPER-2022-031 in preparation}
  {LHCb-PAPER-2022-031 in preparation} \BibitemShut {NoStop}%
\bibitem [{\citenamefont {Yan}\ \emph {et~al.}(2022)\citenamefont {Yan},
  \citenamefont {Peng}, \citenamefont {S\'anchez},\ and\ \citenamefont
  {Pavon~Valderrama}}]{Yan:2022wuz}%
  \BibitemOpen
  \bibfield  {author} {\bibinfo {author} {\bibfnamefont {M.-J.}\ \bibnamefont
  {Yan}}, \bibinfo {author} {\bibfnamefont {F.-Z.}\ \bibnamefont {Peng}},
  \bibinfo {author} {\bibfnamefont {M.~S.}\ \bibnamefont {S\'anchez}}, \ and\
  \bibinfo {author} {\bibfnamefont {M.}~\bibnamefont {Pavon~Valderrama}},\
  }\href@noop {} {\  (\bibinfo {year} {2022})},\ \Eprint
  {http://arxiv.org/abs/2207.11144} {arXiv:2207.11144 [hep-ph]} \BibitemShut
  {NoStop}%
\bibitem [{\citenamefont {Wang}\ and\ \citenamefont
  {Liu}(2022)}]{Wang:2022mxy}%
  \BibitemOpen
  \bibfield  {author} {\bibinfo {author} {\bibfnamefont {F.-L.}\ \bibnamefont
  {Wang}}\ and\ \bibinfo {author} {\bibfnamefont {X.}~\bibnamefont {Liu}},\
  }\href@noop {} {\  (\bibinfo {year} {2022})},\ \Eprint
  {http://arxiv.org/abs/2207.10493} {arXiv:2207.10493 [hep-ph]} \BibitemShut
  {NoStop}%
\bibitem [{\citenamefont {Wang}\ and\ \citenamefont
  {Wang}(2022{\natexlab{b}})}]{Wang:2022neq}%
  \BibitemOpen
  \bibfield  {author} {\bibinfo {author} {\bibfnamefont {X.-W.}\ \bibnamefont
  {Wang}}\ and\ \bibinfo {author} {\bibfnamefont {Z.-G.}\ \bibnamefont
  {Wang}},\ }\href@noop {} {\  (\bibinfo {year} {2022}{\natexlab{b}})},\
  \Eprint {http://arxiv.org/abs/2207.06060} {arXiv:2207.06060 [hep-ph]}
  \BibitemShut {NoStop}%
\bibitem [{\citenamefont {Vijande}\ \emph {et~al.}(2005)\citenamefont
  {Vijande}, \citenamefont {Fernandez},\ and\ \citenamefont
  {Valcarce}}]{Vijande:2004he}%
  \BibitemOpen
  \bibfield  {author} {\bibinfo {author} {\bibfnamefont {J.}~\bibnamefont
  {Vijande}}, \bibinfo {author} {\bibfnamefont {F.}~\bibnamefont {Fernandez}},
  \ and\ \bibinfo {author} {\bibfnamefont {A.}~\bibnamefont {Valcarce}},\
  }\href {\doibase 10.1088/0954-3899/31/5/017} {\bibfield  {journal} {\bibinfo
  {journal} {J. Phys. G}\ }\textbf {\bibinfo {volume} {31}},\ \bibinfo {pages}
  {481} (\bibinfo {year} {2005})},\ \Eprint
  {http://arxiv.org/abs/hep-ph/0411299} {arXiv:hep-ph/0411299} \BibitemShut
  {NoStop}%
\bibitem [{\citenamefont {Garcilazo}\ \emph {et~al.}(2001)\citenamefont
  {Garcilazo}, \citenamefont {Valcarce},\ and\ \citenamefont
  {Fernandez}}]{Garcilazo:2001md}%
  \BibitemOpen
  \bibfield  {author} {\bibinfo {author} {\bibfnamefont {H.}~\bibnamefont
  {Garcilazo}}, \bibinfo {author} {\bibfnamefont {A.}~\bibnamefont {Valcarce}},
  \ and\ \bibinfo {author} {\bibfnamefont {F.}~\bibnamefont {Fernandez}},\
  }\href {\doibase 10.1103/PhysRevC.63.035207} {\bibfield  {journal} {\bibinfo
  {journal} {Phys. Rev. C}\ }\textbf {\bibinfo {volume} {63}},\ \bibinfo
  {pages} {035207} (\bibinfo {year} {2001})}\BibitemShut {NoStop}%
\bibitem [{\citenamefont {Segovia}\ \emph {et~al.}(2013)\citenamefont
  {Segovia}, \citenamefont {Entem}, \citenamefont {Fernandez},\ and\
  \citenamefont {Hernandez}}]{Segovia:2013wma}%
  \BibitemOpen
  \bibfield  {author} {\bibinfo {author} {\bibfnamefont {J.}~\bibnamefont
  {Segovia}}, \bibinfo {author} {\bibfnamefont {D.~R.}\ \bibnamefont {Entem}},
  \bibinfo {author} {\bibfnamefont {F.}~\bibnamefont {Fernandez}}, \ and\
  \bibinfo {author} {\bibfnamefont {E.}~\bibnamefont {Hernandez}},\ }\href
  {\doibase 10.1142/S0218301313300269} {\bibfield  {journal} {\bibinfo
  {journal} {Int. J. Mod. Phys. E}\ }\textbf {\bibinfo {volume} {22}},\
  \bibinfo {pages} {1330026} (\bibinfo {year} {2013})},\ \Eprint
  {http://arxiv.org/abs/1309.6926} {arXiv:1309.6926 [hep-ph]} \BibitemShut
  {NoStop}%
\bibitem [{\citenamefont {Ortega}\ \emph {et~al.}(2013)\citenamefont {Ortega},
  \citenamefont {Entem},\ and\ \citenamefont {Fernandez}}]{Ortega:2012cx}%
  \BibitemOpen
  \bibfield  {author} {\bibinfo {author} {\bibfnamefont {P.~G.}\ \bibnamefont
  {Ortega}}, \bibinfo {author} {\bibfnamefont {D.~R.}\ \bibnamefont {Entem}}, \
  and\ \bibinfo {author} {\bibfnamefont {F.}~\bibnamefont {Fernandez}},\ }\href
  {\doibase 10.1016/j.physletb.2012.12.025} {\bibfield  {journal} {\bibinfo
  {journal} {Phys. Lett. B}\ }\textbf {\bibinfo {volume} {718}},\ \bibinfo
  {pages} {1381} (\bibinfo {year} {2013})},\ \Eprint
  {http://arxiv.org/abs/1210.2633} {arXiv:1210.2633 [hep-ph]} \BibitemShut
  {NoStop}%
\bibitem [{\citenamefont {Ortega}\ \emph {et~al.}(2017)\citenamefont {Ortega},
  \citenamefont {Entem},\ and\ \citenamefont {Fern\'andez}}]{Ortega:2016syt}%
  \BibitemOpen
  \bibfield  {author} {\bibinfo {author} {\bibfnamefont {P.~G.}\ \bibnamefont
  {Ortega}}, \bibinfo {author} {\bibfnamefont {D.~R.}\ \bibnamefont {Entem}}, \
  and\ \bibinfo {author} {\bibfnamefont {F.}~\bibnamefont {Fern\'andez}},\
  }\href {\doibase 10.1016/j.physletb.2016.11.008} {\bibfield  {journal}
  {\bibinfo  {journal} {Phys. Lett. B}\ }\textbf {\bibinfo {volume} {764}},\
  \bibinfo {pages} {207} (\bibinfo {year} {2017})},\ \Eprint
  {http://arxiv.org/abs/1606.06148} {arXiv:1606.06148 [hep-ph]} \BibitemShut
  {NoStop}%
\bibitem [{\citenamefont {Ortega}\ \emph {et~al.}(2014)\citenamefont {Ortega},
  \citenamefont {Entem},\ and\ \citenamefont {Fern\'andez}}]{Ortega:2014fha}%
  \BibitemOpen
  \bibfield  {author} {\bibinfo {author} {\bibfnamefont {P.~G.}\ \bibnamefont
  {Ortega}}, \bibinfo {author} {\bibfnamefont {D.~R.}\ \bibnamefont {Entem}}, \
  and\ \bibinfo {author} {\bibfnamefont {F.}~\bibnamefont {Fern\'andez}},\
  }\href {\doibase 10.1016/j.physletb.2013.12.058} {\bibfield  {journal}
  {\bibinfo  {journal} {Phys. Lett. B}\ }\textbf {\bibinfo {volume} {729}},\
  \bibinfo {pages} {24} (\bibinfo {year} {2014})}\BibitemShut {NoStop}%
\bibitem [{\citenamefont {Diakonov}(2003)}]{Diakonov:2002fq}%
  \BibitemOpen
  \bibfield  {author} {\bibinfo {author} {\bibfnamefont {D.}~\bibnamefont
  {Diakonov}},\ }\href {\doibase 10.1016/S0146-6410(03)90014-7} {\bibfield
  {journal} {\bibinfo  {journal} {Prog. Part. Nucl. Phys.}\ }\textbf {\bibinfo
  {volume} {51}},\ \bibinfo {pages} {173} (\bibinfo {year} {2003})},\ \Eprint
  {http://arxiv.org/abs/hep-ph/0212026} {arXiv:hep-ph/0212026} \BibitemShut
  {NoStop}%
\bibitem [{\citenamefont {Segovia}\ \emph {et~al.}(2008)\citenamefont
  {Segovia}, \citenamefont {Yasser}, \citenamefont {Entem},\ and\ \citenamefont
  {Fernandez}}]{Segovia:2008zz}%
  \BibitemOpen
  \bibfield  {author} {\bibinfo {author} {\bibfnamefont {J.}~\bibnamefont
  {Segovia}}, \bibinfo {author} {\bibfnamefont {A.~M.}\ \bibnamefont {Yasser}},
  \bibinfo {author} {\bibfnamefont {D.~R.}\ \bibnamefont {Entem}}, \ and\
  \bibinfo {author} {\bibfnamefont {F.}~\bibnamefont {Fernandez}},\ }\href
  {\doibase 10.1103/PhysRevD.78.114033} {\bibfield  {journal} {\bibinfo
  {journal} {Phys. Rev. D}\ }\textbf {\bibinfo {volume} {78}},\ \bibinfo
  {pages} {114033} (\bibinfo {year} {2008})}\BibitemShut {NoStop}%
\bibitem [{\citenamefont {Hiyama}\ \emph {et~al.}(2003)\citenamefont {Hiyama},
  \citenamefont {Kino},\ and\ \citenamefont {Kamimura}}]{Hiyama:2003cu}%
  \BibitemOpen
  \bibfield  {author} {\bibinfo {author} {\bibfnamefont {E.}~\bibnamefont
  {Hiyama}}, \bibinfo {author} {\bibfnamefont {Y.}~\bibnamefont {Kino}}, \ and\
  \bibinfo {author} {\bibfnamefont {M.}~\bibnamefont {Kamimura}},\ }\href
  {\doibase 10.1016/S0146-6410(03)90015-9} {\bibfield  {journal} {\bibinfo
  {journal} {Prog. Part. Nucl. Phys.}\ }\textbf {\bibinfo {volume} {51}},\
  \bibinfo {pages} {223} (\bibinfo {year} {2003})}\BibitemShut {NoStop}%
\bibitem [{\citenamefont {Machleidt}(1993)}]{machleidt1993one}%
  \BibitemOpen
  \bibfield  {author} {\bibinfo {author} {\bibfnamefont {R.}~\bibnamefont
  {Machleidt}},\ }in\ \href@noop {} {\emph {\bibinfo {booktitle} {Computational
  Nuclear Physics 2}}}\ (\bibinfo  {publisher} {Springer},\ \bibinfo {year}
  {1993})\ pp.\ \bibinfo {pages} {1--29}\BibitemShut {NoStop}%
\bibitem [{\citenamefont {Broyden}(1965)}]{broyden1965class}%
  \BibitemOpen
  \bibfield  {author} {\bibinfo {author} {\bibfnamefont {C.~G.}\ \bibnamefont
  {Broyden}},\ }\href@noop {} {\bibfield  {journal} {\bibinfo  {journal}
  {Mathematics of computation}\ }\textbf {\bibinfo {volume} {19}},\ \bibinfo
  {pages} {577} (\bibinfo {year} {1965})}\BibitemShut {NoStop}%
\bibitem [{\citenamefont {Azizi}\ \emph {et~al.}(2022)\citenamefont {Azizi},
  \citenamefont {Sarac},\ and\ \citenamefont {Sundu}}]{Azizi:2021pbh}%
  \BibitemOpen
  \bibfield  {author} {\bibinfo {author} {\bibfnamefont {K.}~\bibnamefont
  {Azizi}}, \bibinfo {author} {\bibfnamefont {Y.}~\bibnamefont {Sarac}}, \ and\
  \bibinfo {author} {\bibfnamefont {H.}~\bibnamefont {Sundu}},\ }\href
  {\doibase 10.1140/epjc/s10052-022-10495-7} {\bibfield  {journal} {\bibinfo
  {journal} {Eur. Phys. J. C}\ }\textbf {\bibinfo {volume} {82}},\ \bibinfo
  {pages} {543} (\bibinfo {year} {2022})},\ \Eprint
  {http://arxiv.org/abs/2112.15543} {arXiv:2112.15543 [hep-ph]} \BibitemShut
  {NoStop}%
\bibitem [{\citenamefont {Ferretti}\ and\ \citenamefont
  {Santopinto}(2020)}]{Ferretti:2020ewe}%
  \BibitemOpen
  \bibfield  {author} {\bibinfo {author} {\bibfnamefont {J.}~\bibnamefont
  {Ferretti}}\ and\ \bibinfo {author} {\bibfnamefont {E.}~\bibnamefont
  {Santopinto}},\ }\href {\doibase 10.1007/JHEP04(2020)119} {\bibfield
  {journal} {\bibinfo  {journal} {JHEP}\ }\textbf {\bibinfo {volume} {04}},\
  \bibinfo {pages} {119} (\bibinfo {year} {2020})},\ \Eprint
  {http://arxiv.org/abs/2001.01067} {arXiv:2001.01067 [hep-ph]} \BibitemShut
  {NoStop}%
\bibitem [{\citenamefont {Wang}\ \emph {et~al.}(2021)\citenamefont {Wang},
  \citenamefont {Chen},\ and\ \citenamefont {Liu}}]{Wang:2020bjt}%
  \BibitemOpen
  \bibfield  {author} {\bibinfo {author} {\bibfnamefont {F.-L.}\ \bibnamefont
  {Wang}}, \bibinfo {author} {\bibfnamefont {R.}~\bibnamefont {Chen}}, \ and\
  \bibinfo {author} {\bibfnamefont {X.}~\bibnamefont {Liu}},\ }\href {\doibase
  10.1103/PhysRevD.103.034014} {\bibfield  {journal} {\bibinfo  {journal}
  {Phys. Rev. D}\ }\textbf {\bibinfo {volume} {103}},\ \bibinfo {pages}
  {034014} (\bibinfo {year} {2021})},\ \Eprint
  {http://arxiv.org/abs/2011.14296} {arXiv:2011.14296 [hep-ph]} \BibitemShut
  {NoStop}%
\bibitem [{\citenamefont {Valera}\ \emph {et~al.}(2022)\citenamefont {Valera},
  \citenamefont {Magas},\ and\ \citenamefont {Ramos}}]{Valera:2022elt}%
  \BibitemOpen
  \bibfield  {author} {\bibinfo {author} {\bibfnamefont {J.~A.~M.}\
  \bibnamefont {Valera}, \bibfnamefont {I}}, \bibinfo {author} {\bibfnamefont
  {V.~K.}\ \bibnamefont {Magas}}, \ and\ \bibinfo {author} {\bibfnamefont
  {A.}~\bibnamefont {Ramos}},\ }\href@noop {} {\  (\bibinfo {year} {2022})},\
  \Eprint {http://arxiv.org/abs/2210.02792} {arXiv:2210.02792 [hep-ph]}
  \BibitemShut {NoStop}%
\end{thebibliography}%

\end{document}